\documentclass{elsart}

\usepackage{amsmath,amssymb}
\usepackage{graphicx}
\usepackage{varioref}
\usepackage{slashed}
\usepackage[hyper]{apacite}
\usepackage[bookmarks]{hyperref}

\journal{Studies in History and Philosophy of Modern Physics}

\begin{document}

\begin{frontmatter}
\title{The self-energy of the electron: a quintessential problem in the development of QED}
\author{Jo Bovy}
\ead{jo.bovy@gmail.com}
\address{Centre for Logic and Philosophy of Science, Ghent University, Belgium}










\begin{abstract}
  The development of Quantum Electrodynamics (QED) is sketched from it's
  earliest beginnings until the formulations of 1949, using the
  example of the divergent self-energy of the electron as a
  quintessential problem of the 1930's-40's. The lack of progress
  towards solving this problem led researchers to believe that
  after the conceptual revolution of quantum mechanics a new
  conceptual change was needed. It took a war and a new generation
  of algorithmically inclined physicists to pursue the
  conventional route of regularization and renormalization that
  led to the solution in 1947-1949. Some remarks on contemporary high energy physics are made.\\
\end{abstract}


\end{frontmatter}

\section{Introduction}

The beginning of the twentieth century saw two conceptual
revolutions in physics that would dominate the theoretical physics
of the entire century. The first was Einstein's discovery of
special and general relativity, that abruptly shook our
conceptions of space and time. The lesson of special relativity
was that we can no longer treat space and time as separate
entities, but that they are intermixed and that no treatment can
be totally satisfactory if it does not treat them on an equal
footing. The second conceptual change was the development of
quantum mechanics, that replaced the deterministic framework of
classical physics by an essentially probabilistic one, that only
allows us to calculate the probability of an event. These two
theories were developed independently from each other (though
often by the same researchers). Special relativity was completed
in 1905 by Einstein. This was followed by ten years of labor on
the general theory, which was presented by Einstein in 1916.
Quantum mechanics began in 1901 with the black-body radiation of
Planck, but needed more time to mature. The physicists that were
concerned with it in those early days of the `old quantum theory',
Bohr, Einstein, Planck, were combining classical methods with new
quantization principles and achieved some level of success. But
the true breakthrough in quantum mechanics would only come when a
new, younger generation of physicists stood up, who weren't so
deeply rooted in classical physics any longer, but had grown up
with the new, revolutionary methods of quantum physics. These
physicists like de Broglie, Heisenberg, Dirac and Pauli formulated
quantum mechanics in the modern form in 1925-1927. Unlike Planck,
Bohr and Einstein, who had used quantization methods in their
calculations but had still arrived at exact predictions, the new
theory could only give the probability of the outcome of an
experiment. This conceptual change had remained out of reach for
the old quantum theory and Einstein especially has never come to
terms with it.

But the problems haunting physics were certainly not all resolved
by the developments of the first decades of the twentieth century.
The situation was that there were two theories, one describing the
very fast (special relativity) and one pertaining to the very
small (quantum mechanics). But electrons are very small particles
that sometimes travel at speeds near the speed of light, and
photons, the quanta of light that reappeared in quantum theory
after hundreds of years of a wave picture of light, always travel
at the speed of light. In these cases the two theories had to be
applied both. As a consequence, the interaction of particles with
the electromagnetic field was still poorly understood. This all
constituted a great need to unify the two theories in a consistent
framework. Immediately after the definite formulation of quantum
theory, this new exploration was embarked upon.

The problems were however hard to resolve. Firstly there were two
possible ways to construct the new theory: quantization of the
involved fields or searching for a covariant theory of particles.
The first line of research was the most popular and it started
with de Broglie, and then moved on past the wave mechanics of
Schr\"odinger to the work of Jordan, Pauli and Heisenberg. Their
quantization methods for fields gave rise to new phenomena, such
as the creation and annihilation of particles, which was a new
phenomenon, as the number of particles is a conserved quantity in
ordinary quantum mechanics. Dirac, however, followed the path of
the particles and gave a covariant equation for the behavior of
the electron. One of the consequences thereof was the existence of
anti-particles. Much bigger than these early successes were the
problems of the theory. It was littered with divergences, all
kinds of relevant quantities came out infinite in the theory.

The physicists that worked on quantum electrodynamics in the
1930's were the physicists that had accomplished the big
conceptual breakthrough of quantum mechanics in the 1920's. When
overwhelmed by the problems posed to them by quantum
electrodynamics, they quickly turned away from the methods they
were using and started looking for a new conceptual change. They
were very pessimistic about finding a solution to the problem of
the divergences of the theory in the existing framework and
believed that new conceptual change was needed to surpass the
difficulties.

However, this was not what happened. The second world war started
at the end of the 1930's, which paralyzed research and destroyed
the research infrastructure in Europe. In the meantime a new
generation of American physicists was trained in the American
research laboratories, for whom physics was all about numbers. In
these research facilities the line between experiment and theory
was a fine one, especially when compared to European research
centers. These physicists were all drafted to work in war
laboratories, mostly the Manhattan project and the development of
the Radar, which gave them a strong focus on their problem-solving
and algorithmic capabilities. After the war these young physicists
quickly found the solution of the problems of quantum
electrodynamics. The methods they used were conventional rather
than revolutionary and the expected conceptual change did not take
place. Their solution consisted of regularization and
renormalization and succeeded in discarding the infinities that
had troubled the theory so much. The fact that it took these
physicists to find the solution is found in their training during
the war. Both research paths (field and particle) came to a theory
(found respectively by Julian Schwinger and Richard Feynman) and
not long after, these two approaches were shown to be equivalent
(by Freeman Dyson).

In the following sections of this papers this whole evolution will
be illustrated by reconstructing the development of one of the
quintessential problems of the early quantum electrodynamics, that
of the self-energy of the electron. This is one of the numerous
divergences that haunted the theory from the start and was given
much attention during the subsequent developments. When the final
solution was falling into place both Schwinger and Feynman
considered this problem as one of the first of their new theory
(in their papers \citeNP{Schwinger:1949},
\citeNP{Feynman:spacetime}). We shall see how this problem showed
up after the development of quantum mechanics and what were the
different attempts at solving it. Some attention will go to the
attempts at conceptual change of the 1930's. Then we will follow
the younger physicists during the war and sketch how they came to
the final solution.

In the final section it will be argued that the situation of the
development of quantum electrodynamics is somewhat similar to the
present situation in theoretical high energy physics and this
point will be elaborated upon a bit more.

\section{The global problem: a covariant quantum mechanics}

As has been said before, the problem with quantum mechanics was
that it didn't agree with the principle of relativity of
Einstein's theory, i.e. neither Schr\"odinger's wave mechanics nor
Heisenberg's matrix mechanics was covariant. This is immediately
clear when we consider the Schr\"odinger equation for a mechanical
particle without spin
\begin{equation}
  i \hbar \frac{\partial}{\partial t} \psi = H \psi \quad ; \quad
  \psi \in \mathcal{H} = L^2(\mathbb{R}^3) \ ,
\end{equation}
here the Hamiltonian $H$ is
\begin{equation}
  H = -\frac{\hbar^2}{2m}\Delta + V(x) \ .
\end{equation}
This gives a differential equation of first order in the time
derivative and of second order in the place derivative, which
can't be Lorentzcovariant, as space and time have to be treated
equally in special relativity. A covariant version of the
Schr\"odinger equation had to be found.

\subsection{The first attempt: the Klein-Gordon equation}

The first covariant quantum mechanical equation was found by using
the correspondence principle of quantum mechanics together with
the new insights relativity had brought. The correspondence
principle is a way to quantize classical equations, by changing
space coordinates $x_k$ by multiplication operators on the state
space $\mathcal{H} = L^2(\mathbb{R}^3)$. On a wave function this
gives $\psi (x) \rightarrow x_k \psi (x)$. The classical momenta
$p_k$, however, are replaced by differential operators: $\psi (x)
\rightarrow \frac{\hbar}{i} \frac{\partial}{\partial x_k}
\psi(x)$. The classical energy is substituted by the
energy-differential operator $i\hbar \frac{\partial}{\partial t}$.
By making these changes, one becomes the Schr\"odinger equation
out of the classical expression for the energy of a particle ($E =
\frac{p^2}{2m} + V(x)$), as one can readily check.

In special relativity space and time coordinates are considered
together and they form covariant vectors $(x_\mu) =
(x_0,x_1,x_2,x_3)$ and $(p_\mu) = (p_0,p_1,p_2,p_3)$ (here $x_0 =
ct$ and $p_0 = E/c$). The correspondence principle connects these
two vectors
\begin{equation}
  p_\mu = i \hbar \frac{\partial}{\partial x^\mu}
\end{equation}
which gives again
\begin{align}
  p_0 & = \frac{E}{c} = i \hbar \frac{\partial}{c \partial t} \\
  p_k & = \frac{\hbar}{i} \frac{\partial}{\partial x_k}  \ .
\end{align}
In special relativity the expression for the energy of a free
particle is given by
\begin{equation}\label{RelEn}
  E^2 = m_0^2 c^4 + p^2 c^2 \ .
\end{equation}
This gives
\begin{equation}
  m_0^2 c^2 = \frac{E^2}{c^2} - p \cdot p = p^\mu p_\mu
\end{equation}
(Einstein summation convention). By quantizing this equation we
find the simplest covariant quantum mechanical equation, the
Klein-Gordon equation for a free particle with spin zero:
\begin{equation}
  p^\mu p_\mu \psi = m_0^2 c^2 \psi \quad , \psi \in \mathcal{H}
\end{equation}
or by substitution
\begin{equation}
  \Big( \square + \big(\frac{m_0 c}{\hbar}\big)^2\Big) \psi = 0
\end{equation}
with the d'Alembertian given by
\begin{equation}
  \square = \partial^\mu \partial_\mu = \frac{\partial^2}{c^2 \partial
  t^2}- \sum_{k=1}^3{\frac{\partial^2}{\partial x_k^2}} \ .
\end{equation}

So we see that as the Schr\"odinger equation is the quantization
of the classical expression for the energy of a particle, the
Klein-Gordon equation is the quantization of the relativistic
expression for the energy. When we solve this Klein-Gordon
equation, we find solutions with positive energies as well as
solutions with negative energies. This shouldn't be surprising,
because the relativistic expression for the energy of a particle
(\ref{RelEn}) has always positive and negative solutions. In
quantum mechanics these negative solutions give interpretation
difficulties (non-positiveness of the probability densities).

\subsection{Staring into the fireplace: the Dirac equation}

One night Dirac was staring into a fire when he suddenly thought
he wanted a relativistic wave equation that was linear in the
space-time derivatives $\partial_\mu \equiv
\frac{\partial}{\partial x^\mu}$. This equation had to have the
form of `some linear combination of the $\partial_\mu$'s working
on a field $\psi$, that is equal to a constant times that field'.
When we write this down, this becomes
\begin{equation}\label{Dirac}
    (i \gamma^\mu \partial _\mu - m)\psi = 0 \ .
\end{equation}
If the $\gamma^\mu$'s  were numbers, the vector $\gamma^\mu$ would
define a direction in space-time and this would break the
covariance of the theory. This is why the $\gamma^\mu$'s can't be
numbers. They were identified as matrices, subjected to the
anti-commutation relation
\begin{equation}
  \{ \gamma^\mu, \gamma^\nu \} = 2 \eta^{\mu \nu} \ ,
\end{equation}
with $\eta^{\mu \nu}$ the Minkowski metric of special relativity.
The wonderful thing about this equation is that it describes the
spin property of electrons, which is a purely quantum mechanical
effect. This is the equation for the behavior of spin-1/2
particles.

The Dirac equation, just like the Klein-Gordon equation, has
negative energy solutions that cause problems. Dirac then proposed
to consider as the vacuum state that state in which all negative
energy states are filled. Then the Pauli exclusion principle will
force any extra electron to take a positive energy state. This
vacuum state is called the \emph{sea of electrons}. When an
electron is excited from a negative energy state to a positive
energy state, it leaves a hole in the sea behind. Dirac proposed
that this hole be considered a particle itself, with a positive
charge now. First he identified this particle with the proton, but
subsequent developments showed that this couldn't be the case. He
then proposed that it was a new particle, identical to the
electron but for its charge, which was $+e$. He called it the
\emph{positron}. The positron was experimentally observed in 1932
by Carl D. Anderson. This way of looking at the sea of electrons
allows us to view pair creation as an electron jumping from a
negative energy state to a positive energy state, leaving a hole,
i.e. a positron, behind. This theory became known as Dirac's
\emph{hole theory}.

\section{The problem: the divergence of the self-energy of the
electron}

The self-energy is the energy that an electron in free space,
isolated from other particles, fields, or lightquanta, possesses.
In the classical theory it posed no problem, but after the
development of quantum theory, it became a critical problem for
theoretical physics. The problem was first noted by Pauli and
Heisenberg in their papers \citeNP{HeisenbergPauli},
\citeNP{HeisenbergPauli2}. In \citeNP{Weisskopf:1939} Weisskopf
gives a good review of the problem. The self-energy of the
electron is given by
\begin{equation}
  W = T + (1/ 8 \pi) \int (E^2 + H^2) \mbox{d}V \ ,
\end{equation}
with $T$ the kinetic energy of the electron (which for a
non-moving electron is just equal to the rest-energy $mc^2$) and
$E$ and $H$ the electric and magnetic field strengths. d$V$ is the
volume-element.

In classical electromagnetism, the electric field $E$ of a free
electron is equal to $e/r^2$, $r$ is the distance to the electron.
When we assume that the electron does not have any spin, the
magnetic field $H$ equals zero. The self-energy is then given by
\begin{equation}
  W \sim mc^2 + \int \frac{e^2}{r^4} \mbox{d}V \sim mc^2 + \int
  \frac{e^2}{r^2}\mbox{d}r \ .
\end{equation}
If the radius of the electron were zero, this integral would run
from zero to infinity and thus constitute a linear
divergence\footnote{Here it is said that the integral
$\int_0^\infty \mbox{d}r/r^2$ is linearly divergent. The reason we
speak of a linear divergence follows from the transformation $r
\rightarrow \frac{1}{R}$. We then have
\begin{align*}
    \mbox{d}r &\rightarrow - \mbox{d}R/R^2 \\
    0 &\rightarrow \infty\\
    \infty &\rightarrow 0 \,
\end{align*}which transforms the integral into \[\int_\infty^0  -
\frac{\mbox{d}R}{R^2} R^2 = \int_0^\infty \mbox{d}R \ .\] This is
clearly linearly divergent.}. When we assume that the electron has
a radius equal to $a$, the integral is calculated to be
$\int_a^\infty e^2 \mbox{d}r/r^2 = e^2/a$. This is the reason why
classically it was assumed that the electron has a finite radius,
and thus isn't a point particle.

The development of quantum mechanics made the self-energy,
however, into a critical problem. We can discern three reasons for
this:\\
(a) Quantum mechanics shows that the radius of the electron has to
be zero, i.e. that the electron is a point particle. This is
because we can prove that the product of the charge densities in
two different points equals a delta-function, i.e. a function that
peaks in one place and is equal to zero everywhere else. For a
free electron this means that the probability that we find charge
densities in two different places equals zero. Thus the charge has
to be concentrated in one point. Like we saw in the last paragraph
this means that the contribution of the electrostatic energy
diverges. This divergence is linear.\\
(b) Relativistic quantum mechanics showed that the electron
possesses an intrinsic spin (which was first experimentally
observed by Stern and Gerlach). Because of this intrinsic
spin-property of the electron, the value of $H$ will not be equal
to zero anymore, as the spin induces a magnetic field and an
alternating electric field. These contributions to the self-energy
thus have to be added to that of the electrostatic energy.\\
(c) Finally, quantum mechanics of the electromagnetic field
postulates the existence of field strength fluctuations in free
space. The divergence of the self-energy as a consequence of these
fluctuations is bigger than that of the electrostatic energy. The
energy of the fluctuations is $W_{\mbox{fluct}} \sim e^2 h/mca^2$
for an electron of radius $a$. This constitutes a quadratic
divergence.

\section{The self-energy in Dirac's one-particle theory}

The first calculation of the self-energy of the electron was
performed using Dirac's one-particle theory. This is the theory
that uses the Dirac equation, but doesn't use the vacuum state
with all negative energy states filled, as does Dirac's hole
theory. The calculation uses standard quantum mechanical
perturbation theory to find the $e^2$-contribution to the
self-energy. The whole Hamiltonian of the system is split into the
Hamiltonian of the unperturbed system $H_0$ (here the Hamiltonian
of the free electron and the Hamiltonian of the free
electromagnetic field) and the interaction Hamiltonian $H'$ (here
the interaction of the electron with the electric field). It is
assumed that the contribution of the interaction Hamiltonian is
small compared to the contributions of the unperturbed system. The
Hamiltonian of Dirac's one-particle theory is given by
\begin{align}
  H & = H_0 + H'\\
    & = \sum \hbar k c
    a_{k\mathbf{\epsilon}}^*a_{k\mathbf{\epsilon}} +
    H_{\mbox{Coulomb}} + \beta m + \alpha \cdot
    (\mathbf{p} - e \mathbf{A})
\end{align}
with the interaction term linear in A:
\begin{equation}
  H' = - e \alpha \cdot \mathbf{A} \ ,
\end{equation}
with $\mathbf{A}$ the vector potential of electromagnetism and
$\alpha$ satisfying
\begin{equation}
  \alpha_i \alpha_j + \alpha_j \alpha_i = \delta_{ij} \ .
\end{equation}

The first contribution to the self-energy of the electron is then
given by
\begin{equation}
  \Delta W = \sum{\frac{|<n|H'|0>|^2}{E_0-E_n}} \ ,
\end{equation}
from standard perturbation theory. $|n>$ are the unperturbed
states of $H_0$. We have
\begin{equation}
  E_0-E_n = m \mp \sqrt{m^2+k^2} - k \ ,
\end{equation}
which for large $k$ goes like $m$. The matrix element from the
numerator is for large $k$
\begin{equation}
  |<n|H'|0>|^2 \sim \frac{e^2}{k} \ .
\end{equation}
Since we can replace the sum by an integral
\begin{equation}
  \sum_n \rightarrow \int \mbox{d}^3k \sim \int_0^\infty k^2
  \mbox{d}k \ ,
\end{equation}
we find
\begin{equation}
  \Delta W \sim \int \frac{k\mbox{d}k}{m} \ ,
\end{equation}
which is a quadratic divergence. This situation is far worse than
that in the classical theory, where, as we saw, we only have a
linear divergence of the self-energy. \citeNP{Waller} was the
first to find this result. \citeNP{Oppenheimer} and
\citeNP{Rosenfeld} came to similar conclusions.

\begin{figure}
  \begin{center}
    \includegraphics[width=0.5\textwidth]{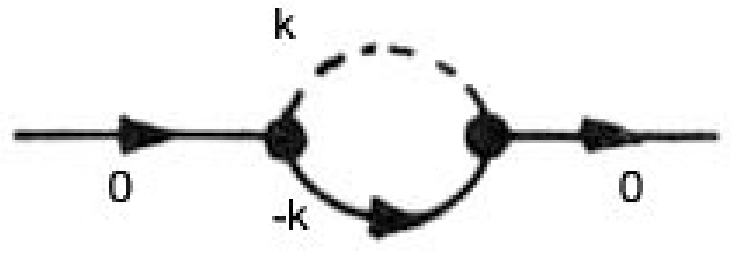}
    \caption[Self-energy of the electron to first order]{The
    self-energy of the electron to first order. On the left the
    electron is at rest. Then a virtual photon forms. On the right,
     the electron is at rest again.}\label{Disp}
  \end{center}
\end{figure}

In figure \vref{Disp} we see this situation graphically. On the
left we see a free electron at rest. A virtual photon appears,
which forces the electron to have a momentum equal to $-k$,
because of the conservation of momentum. This photon disappears
and the electron is at rest again.

\section{Weisskopf's calculation of the self-energy in Dirac's
hole theory}

\subsection{1934: calculation of the $e^2$ contribution}

\citeNP{Weisskopf:1934}b used Dirac's hole theory to calculate the
self-energy of the electron to first order . He divided the total
self-energy in an electrostatic part $E^S$ and an electrodynamic
part $E^D$. He was able to show that the contribution of the
electrostatic part only constituted a logarithmic divergence,
which was better than what we just found using one-particle
theory. For the electrodynamic part, however, he found, just as in
one-particle theory, a quadratic divergence.

Not much later he received a letter from Furry, who reported to
him that he had done the calculation of the self-energy in hole
theory himself and had only found a logarithmic divergence for the
electrodynamic part. Weisskopf readily admitted that he had made a
mistake and published a correction \citeNP{Weisskopf:correction}a.
Let's recall the most important results found in this paper. One
particle theory led to
\begin{align}
  E^S & = \frac{c^2}{h}\int_0^\infty \mbox{d}k + \mbox{finite terms}\\
  E^D & = \frac{e^2}{h}\Big[\frac{m^2c^2}{p(m^2c^2+p^2)^{1/2}}\log
  \frac{(m^2c^2+p^2)^{1/2}+p}{(m^2c^2+p^2)^{1/2}-p}-2\Big]\int_0^\infty
  \mbox{d}k \nonumber\\
  & \ \ +\frac{2e^2}{h(m^2c^2+p^2)^{1/2}}\int_0^\infty
  k \mbox{d}k \ .
\end{align}
The divergence of the total self-energy is thus quadratic. In hole
theory the following expressions were found
\begin{align}\label{Weiss34}
  E^S & =
  \frac{e^2}{h(m^2c^2+p^2)^{1/2}}(2m^2c^2+p^2)\int_{k_0}^\infty
  \frac{\mbox{d}k}{k} + \mbox{finite terms}\\
  E^D & = \frac{e^2}{h(m^2c^2+p^2)^{1/2}}(m^2c^2-\frac{4}{3}p^2)\int_{k_0}^\infty
  \frac{\mbox{d}k}{k} + \mbox{finite terms} \ \label{Wei}.
\end{align}
This is only a logarithmic divergence of the $e^2$ term.

These results, however, could not be satisfactory. Heisenberg
repeated the calculation himself and found the same results, but
in a letter to Weisskopf he characterized this solution as
``implausible and suspicious'' (cited in \citeNP{Schweber}). The
reason for this was that because of relativistic invariance
reasons, one would expect an expression of the following form
\begin{equation}
  E^S + E^D = \mbox{constant} \frac{e^2}{h} \sqrt{m^2c^2 + p^2}
  \int \frac{\mbox{d}k}{k} \ ,
\end{equation}
as the expression $m^2c^2 + p^2$ is relativistically invariant,
whereas the sum of the expressions in (\ref{Weiss34}-\ref{Wei}) is
not. The lack of correct relativistic invariance was the biggest
problem in pre-war quantum electrodynamics.

\subsection{1939: logarithmic divergence of the full self-energy}

The most extensive treatment of the problem of the self-energy of
the electron before the second world war was given in 1939 by
Weisskopf. In \citeNP{Weisskopf:1939} he argued that the
self-energy is logarithmic divergent in every order in hole
theory. This was a major breakthrough. In 1936 Dirac, Heisenberg
and Weisskopf had solved the problem of the vacuum
polarization\footnote{The problem of the vacuum polarization was
another problem that arose after the development of quantum
mechanics, much in the same way as the self-energy of the
electron. It is also hampered by divergences. The problem exists
because virtual particle/anti-particle pairs in the vacuum, when
charged, constitute an electric dipole. A electromagnetic field
orientates these dipoles.} by using relativistically invariant
subtraction principles and charge renormalization. Kramers, Pauli
and Fierz had, in addition, already given a procedure to get rid
of logarithmic divergences in 1937-1938. Weisskopf even went as
far as to claim that all divergence problems in quantum
electrodynamics could be solved by using these principles
\cite{Weisskopf:1936}.

But these new insights were never brought together to construct a
diver-\\gence-free hole theory, not even up to first order. This
was most probable because the physics community in the thirties
did not believe that this was the right way to go and because
serious questions were being asked about the methods to get rid of
the infinities. As a reaction to a paper of \citeNP{Dirac:1934}
that introduced methods to solve problems with infinities,
\citeNP{Peierls} questioned the uniqueness of the methods to
subtract infinities, precisely because they are infinite. Pauli
reacted shocked to a similar attempt by Heisenberg. Pauli noted
that he didn't believe that these methods to get rid of infinities
could lead to results that were not already known\footnote{See
\citeNP{Miller} for a more extensive treatment of this
substraction physics and the reactions it led to.}.

The first attempts to come to a renormalization procedure were
thus not met with a lot of enthusiasm by the European physics
community. A unified theory of radiation remained an open problem.

\section{Looking for conceptual change}

All the attempts in the 1930's to solve the problem of the
divergences failed. This had much to do with the pessimism of the
leading figures in physics. Bohr, Pauli, Dirac and Heisenberg
didn't see how the problem could be solved in the framework of
quantum mechanics. They thought that the solution would come from
new concepts, in the same manner as the problems of the old
quantum theory were solved by introducing a whole new framework,
i.e. the probabilistic quantum mechanics. As early as 1930 in an
article concerning the self-energy of the electron,
\citeNP{Heisenberg:1930} remarked that the problem of the
divergence of the self-energy does not appear in a lattice-world,
i.e. a discrete model of space-time. He doesn't elaborate on it
any further there, because he immediately sees the difficulties of
this proposal, most notably that a lattice breaks relativistic
invariance.

In 1938 Heisenberg is still defending conceptual renewal. In
\citeNP{Heisenberg:1938} he claims that after the light speed $c$,
that became a fundamental unit in relativity theory, and after
Planck's constant, which rose to prominence in the quantum
mechanical revolution, the time has come for a new fundamental
constant, a fundamental length now, that would delineate the area
of validity for the classical theories, where the theory of fields
and particles can be applied without difficulties, and below which
new phenomena will appear. He regards the self-energy of the
electron as one of the reasons for such a fundamental length.

Many other alternative formalisms were developed at the end of the
1930's. Wheeler suggested that the formalism of state vectors and
quantum fields should be replaced by a formalism based on
observables only, such as the $S$-matrix he introduced in 1937.
This $S$-matrix contains scattering amplitudes (the $S$-matrix
became a vital part of modern quantum mechanics later, and Feynman
gave easy rules to calculate its elements up to any desired order
in perturbation theory, see \emph{infra}). Wheeler and Feynman
also worked on a formalism that sought to eliminate the
electromagnetic field, by deriving all electromagnetic properties
by an interaction at a distance. Dirac, radically, suggested the
use of states of negative probability.

This looking for conceptual change held the clear development of a
renormalization theory for divergences back, when all the
ingredients to found it were already available. When the second
world war broke out, the research was freezed and when it was
finally over, the research climate had changed considerably. The
center of post-war physics was no longer situated in Europe, but
had moved to the United States, and the solution that hadn't been
possible in pre-war Europe, was found by a new generation of
physicists.

\section{Physicists in the second world war}

The second world war brought great changes with it. European
research was completely paralyzed and lots of European physicists
emigrated to the United States (viz. Einstein, Wigner). Many
physicists were set to work in the war effort. This included
projects like the Manhattan project and the development of the
Radar. Many of the developments in later quantum electrodynamics
came out of the efforts of physicists that had worked in those war
laboratories. The head of the Manhattan project was Oppenheimer
and leading the theoretical division of the Los Alamos laboratory
was Hans Bethe, who would play an important part in the post-war
developments. At the Los Alamos laboratory we also find a young
Richard Feynman. He worked in the computation facilities and he
designed methods to do the great amounts of calculations that were
necessary for the development of the nuclear bomb.

At the radiation laboratory at MIT we find Julian Schwinger during
the war, doing theoretical work concerning the radar. The insights
he gained there about radiation, he would apply to quantum
mechanics after the war. Freeman Dyson was working as an analyst
for a British fighter plane division.

We see that a lot of theoretical physicists were working on
practical applications of their work during the war. This work was
focussed on problem-solving and very quantitative by nature. This
way they developed crucial skills for solving the problems of
quantum electrodynamics. The problem-solving and algorithmic
nature of the work of Feynman at Los Alamos, would show in his
subsequent work on quantum electrodynamics. He would give a simple
algorithm to solve problems in it.

This emigration of prominent physicists to the United States and
especially the experience the young American physicists gained
during the war, shifted the center of scientific research after
the world war to the United States. Schwinger, Feynman and Dyson
(and Tomonaga in Japan) would solve the problems that had troubled
physicists during the 1930's in two different ways, which turned
out to be equivalent. Thus they found the most accurate theory
known as yet.

\section{1947-1950: Renormalization}

\begin{figure}
  \begin{center}
    \includegraphics[width=0.6\textwidth]{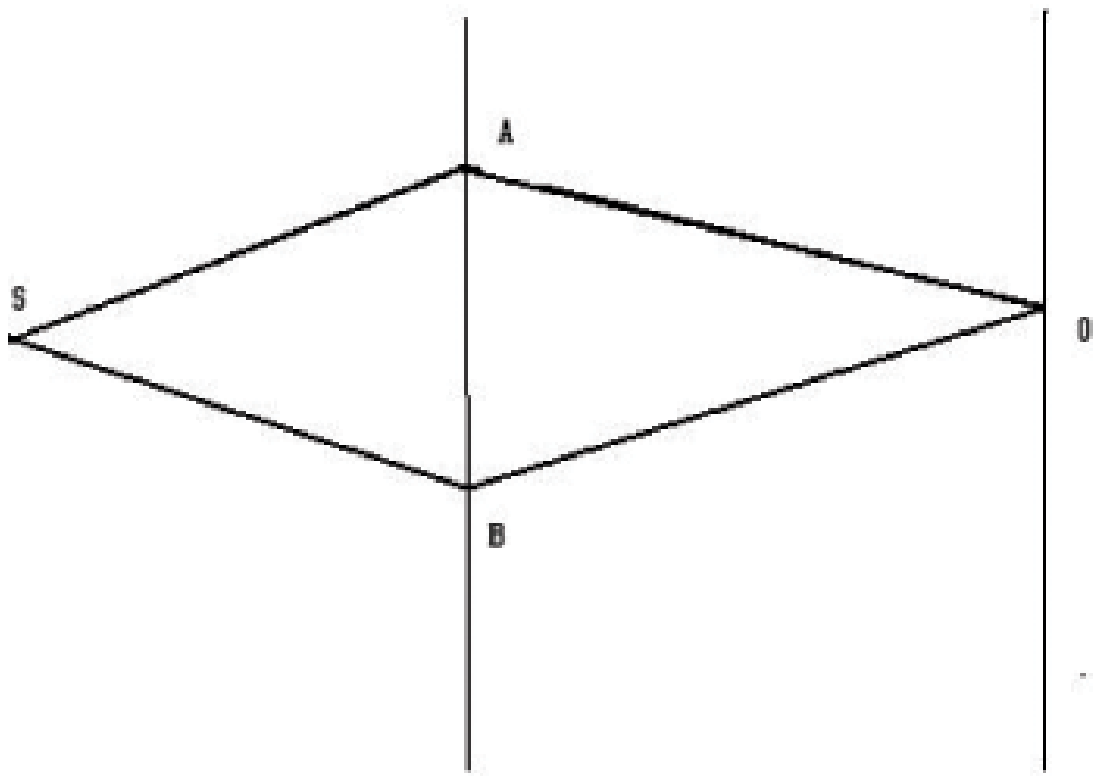}
    \caption{The two slit experiment.}\label{Twoslits}
  \end{center}
\end{figure}

During the years 1947-1950 Schwinger and Feynman both found a
formalism which transformed quantum electrodynamics into a sound,
divergence-free theory. The method of Schwinger was tedious and
complicated, whereas Feynman's gave a simple method to solve
problems concerning radiation. We shall succinctly describe
Feynman's formalism and see how he uses it to work on the
self-energy of the electron. It all begins with a reformulation of
quantum mechanics.

\subsection{Feynman's path-integral formulation of quantum
mechanics}

We can illustrate the reformulation that Feynman gave of the
quantum mechanics of Schr\"odinger, Heisenberg and Dirac by way of
a familiar example. One of the most famous experiments in quantum
mechanics is certainly the double-slit experiment. This experiment
was first performed by Young for the case of light waves. These
were sent to a screen in which two slits were made, and observed
on a screen situated behind this first screen (see figure
\vref{Twoslits}). The observation Young made was that these two
rays interfered and this was thought to be a sufficient proof for
the wave character of light. In quantum mechanics, however, we
have wave-particle duality, such that when we sent particles to a
screen with two slits, we will also find an interference pattern
on the observation screen. This is because we have to add the
amplitudes of the two electron paths and square them to get the
probability (in Born's interpretation of the Schr\"odinger
equation). So in figure \ref{Twoslits} we have two paths, we
simply add the amplitudes of these two paths and square this sum.

\begin{figure}
  \begin{center}
    \includegraphics[width=0.6\textwidth]{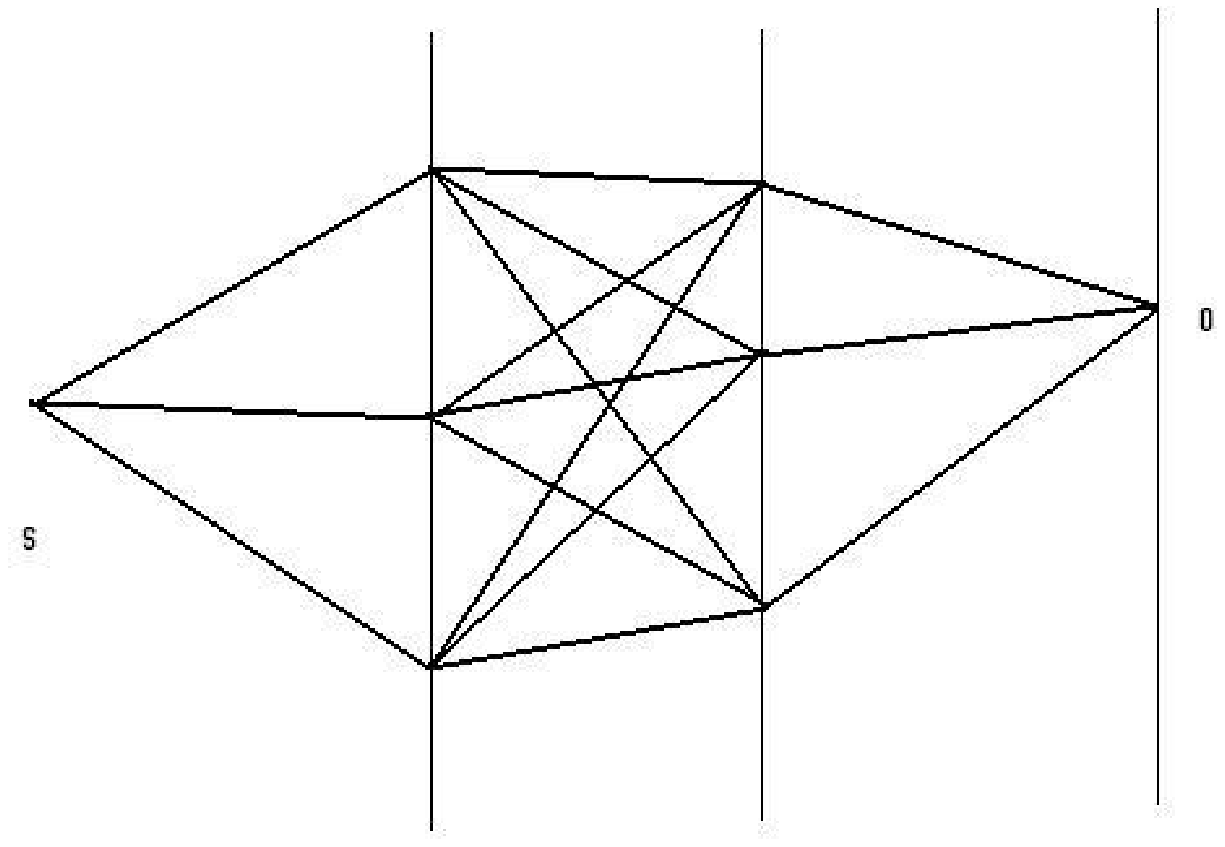}
    \caption{Various possibilities.}\label{slits}
  \end{center}
\end{figure}

\begin{figure}
  \begin{center}
    \includegraphics[width=0.6\textwidth]{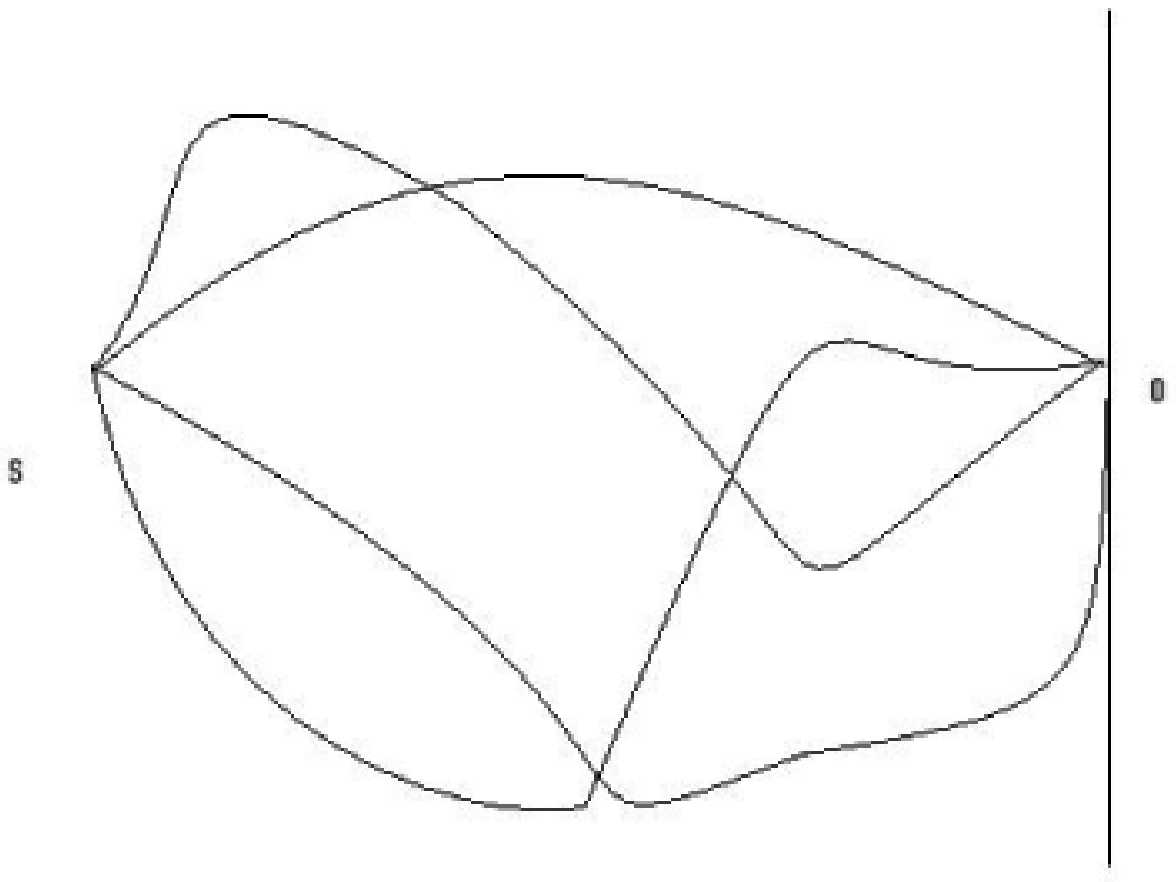}
    \caption{All possible paths.}\label{Feyn}
  \end{center}
\end{figure}

We can now ask what happens when we drill an extra hole in the
middle screen. The solution is simple: we add the amplitude of the
third path to the sum of the first two and square this expression
to get the probability. We can also ask what happens when we add
another screen, with a couple of holes drilled in it. This gives
us the situation of figure \vref{slits}. Now we have to add the
amplitudes of all the possible paths shown and square this whole
expression. This process can go on. When we add a fourth, a fifth
screen, when we drill a fourth, a fifth hole in each screen, we
always have to consider all possible paths and add amplitudes.
When we add an infinity of screens, the whole space will be
filled, when we drill an infinity of holes in a screen, the screen
will disappear. This way we finally see that we have to add the
amplitudes of \emph{all possible paths between s and o} that the
particle can take (see figure \vref{Feyn}). This is the
possible-paths interpretation of quantum mechanics of Feynman.

Of course, this has to be formalized, and when we do this we get
an integral over all possible paths, with the amplitude of a path
given by $e^{iS}$, with $S$ the classical action of the path,
which is given by an integral over the kinetic energy minus the
potential energy of the particle (this last sum is the Lagrangian
$L$). We find that the probability is given by the expression
\begin{equation}\label{IntF}
  \int{Dq(t) e^{i \int_0^T{dt L(\dot{q},q)}}}
\end{equation}

\subsection{The self-energy of the electron in Feynman's QED}

When we use this formulation and replace the classical action by a
relativistic invariant one for fields, we become a quantum field
theory. We can do this for fields described by the Klein-Gordon
equation as well as for fields described by the Dirac equation.
The problem then reduces to calculating the integral (\ref{IntF}).
It is, however, not possible to calculate this integral exactly,
so it has to be calculated approximately, by expanding it in
orders of the coupling constants (for quantum electrodynamics this
is the fine-structure constant $\alpha = \frac{e^2}{\hbar c}
\approx 1/137$). Feynman gave simple rules to construct the terms
in this expansion for a given problem. One simply draws the
situation one would like to calculate, for example the $e^2$
contribution leads to the diagram \vref{Disp}, and then the
Feynman rules tell us how to associate expressions with these
particles, lines and nodes. For the matrix-element of the $e^2$
contribution to the self-energy we get the following, complicated
expression
\begin{align}
  &e^2\int{\frac{\mbox{d}^4 k}{(2 \pi)^4}} \frac{\bar{u}^{r'}(p)}{(2 \pi)^{3/2}}
  \Bigg(
  \frac{m}{p_0}\Bigg)^{1/2} \cdot \frac{i}{(2 \pi)^4}\frac{\slashed{p}-\slashed{k} + m}{(p-k)^2-m^2+ i\varepsilon}
  \cdot \frac{-i g_{\mu \nu}}{k^2 + i \varepsilon}\nonumber \\ &\Bigg(\frac{1}{2
  \pi}\Bigg)^4 i^2 \gamma_\mu \gamma_\nu \cdot \frac{u^r(p)}{(2
  \pi)^{3/2}}\Bigg( \frac{m}{p_0}\Bigg)^{1/2} \ ,
\end{align}
but what is important is that this behaves as
\begin{equation}
  \int{\mbox{d}^4k \frac{\slashed{k}}{k^4}}
\end{equation}
for large $k$.

Thus it seems that we have a linear divergence. Because of the
symmetry of the integral, however, this contribution vanishes,
which leads to a logarithmic divergence.

\citeNP{Feynman:spacetime} is able to show, using this formalism,
that the first order correction to the self-energy of the electron
is finite. He uses a modification of quantum electrodynamics, in
which the Dirac-function that appears in his expression, is
replaced by a function of small width and great height (that thus
approximates a Dirac-function\footnote{A Dirac-function is not a
function in the mathematical sense of the word. It is actually a
\emph{distribution}, but is generally referred to by physicists as
a function. The Dirac-distribution can be approximated by a
variety of mathematical functions.}). This way he avoids the
divergence coming from this singularity.

As his solution is long and tedious, it is perhaps more
instructive to give some cursory remarks regarding the process of
regularization and renormalization. For example, we can solve a
logarithmic divergence by this procedure as follows.
Regularization states that one should not expect this theory to
hold to arbitrarily high energies, the theory only holds up to
some value for the energy, say $\Lambda$. This transforms the
integral to
\begin{equation}
  \int^\Lambda{\mbox{d}^4k \frac{1}{k^4}} \ .
\end{equation}
We can interpret this $\Lambda$ as the value at which the
expansion is not valid anymore. At the value $\Lambda$ the second
term in the expansion becomes as large as the first term. Then one
can not claim any longer that this expansion will give accurate
results. This drops the infinities, but we do introduce a new,
unknown constant $\Lambda$.

Here comes renormalization into play. What we measure are of
course physical quantities. When we couple theory to experiment,
we have to express the theoretical results in terms of the
physical couplings. When we do this we see that the constant
$\Lambda$ vanishes from the expressions.

For instance, using the example of vacuum polarization, we see
that this polarization effects the charge density of the electron.
That way, it changes the charge of the electron, because the
charge is the integral over the charge density. As a consequence,
the charge of the electron will not longer equal $e$. When we now
want to express results of vacuum polarization, we shouldn't use
$e$, which is the theoretical coupling, but the physical coupling,
i.e. the changed value of the charge. Expressing the theoretical
results by using the physical couplings rather than the
theoretical ones, produces finite results (since the infinities
cancel in the process of rewriting the expressions in terms of the
physical couplings).

\section{Conclusion}

We have seen how applying quantum mechanics to the interaction
between fields and particles led to problems that were apparently
out of reach of the quantum theory. The theory that was developed
in the thirties of the previous century, was haunted by a lot of
divergences, that couldn't be of a physical nature.  The European
physicists had no trust in the methods used up to then and thought
that only a new conceptual change could lead to a solution of
these problems. It took a new generation of American physicists,
trained with a strong emphasis on the algorithmic and
problem-solving aspects of physics, to see that the solution could
be found in a conventional manner, by disposing of the divergences
by renormalization. This way they developed quantum
electrodynamics, and more generally, quantum field theory.

In the 1950's and 1960's quantum field theory was developed
further and applied to the other forces of nature. Using gauge
theories a unification of all known forces except gravity was
achieved. Weinberg, Salam and Glashow unified the electromagnetic
force with the weak nuclear force using quantum field theory. The
strong nuclear force was described as a quantum field theory in
quantum chromodynamics (QCD). This eventually led to the grand
unification theories (GUT's) that unified these three forces in
one consistent framework.

The only remaining force that eluded the quantum mechanical
framework was gravity. A quantum field theory for gravity can not
be renormalizable. Emphasis in the last twenty years of the
twentieth century in high-energy physics was on the problem of
finding this quantum description and unifying all forces in one
theory. Highest hopes in the theoretical physics community rest on
string theory. This theory is founded on the principles of quantum
field theory, but the basic entities aren't particles anymore, but
one-dimensional strings. It is hoped that this field theory will
give a consistent quantum description of gravity. String theory
has as virtues that it naturally contains a spin-2 particle, a
particle that can function as the graviton, the messenger-particle
of gravity and that it provides a framework to unify all four
forces of nature. String theory, however, has still a lot of
problems. For instance, it predicts that space-time has ten
dimensions, instead of the usual four. The search for the solution
of these problems already takes up a few decades.

In certain aspects, the situation we see today is similar to the
situation that confronted physicists in the 1930's. The problem
then was unifying the theory of special relativity with the
quantum theory, and the problem today is unifying the theory of
general relativity with the quantum theory. The problems then, the
divergences, seemed unsolvable, and the problems today again make
the physical interpretation of the theory difficult, e.g. the
extra dimensions. The manner in which a solution is sought,
however, is completely different. The leading physicists of the
1930's had brought about the conceptual change of quantum theory
and believed that only a new conceptual change could lead to
satisfying results. We saw that this judgement was mistaken. The
physicists that started the research for a quantum mechanical
description of gravity in the second half of the twentieth
century, were physicists who had accomplished many successes with
quantum field theory. It was only natural for them to look for a
solution using the framework of quantum field theory, which led
them to string theory. However, now that we see that this theory
does not arrive at a solution, we may ask whether it could not be
that this time we do need a conceptual change.

\section*{Acknowledgements}

I would like to thank Prof. Joke Meheus of Ghent University for
helpful comments on a first draft.

\nocite{*}
\bibliographystyle{apacite}

\end{document}